\documentclass[thmsa]{article}
\usepackage{sw20lart}

\input{tcilatex}
\begin{document}

\title{Yet another resolution of the Gibbs paradox: an information theory approach%
\thanks{%
Presented at MaxEnt 2001, the 21th International Workshop on Bayesian
Inference and Maximum Entropy Methods (August 4-9, 2001, Baltimore, MD, USA).%
}}
\author{Chih-Yuan Tseng\-\thanks{%
E-mail address: ct7663@csc.albany.edu} and Ariel Caticha\thanks{%
E-mail address: Ariel.Caticha@albany.edu} \\
{\small Department of Physics, University at Albany-SUNY, }\\
{\small Albany, NY 12222, USA.}}
\date{}
\maketitle

\begin{abstract}
The ``Gibbs Paradox'' refers to several related questions concerning entropy
in thermodynamics and statistical mechanics: whether it is an extensive
quantity or not, how it changes when identical particles are mixed, and the
proper way to count states in systems of identical particles. Several
authors have recognized that the paradox is resolved once it is realized
that there is no such thing as \emph{the} entropy of a system, that there
are \emph{many} entropies, and that the choice between treating particles as
being distinguishable or not depends on the resolution of the experiment.
The purpose of this note is essentially pedagogical; we add to their
analysis by examining the paradox from the point of view of information
theory. Our argument is based on that `grouping' property of entropy that
Shannon recognized, by including it among his axioms, as an essential
requirement on any measure of information. Not only does it provide the
right connection between different entropies but, in addition, it draws our
attention to the obvious fact that addressing issues of distinguishability
and of counting states requires a clear idea about what precisely do we mean
by a state.
\end{abstract}

\section{Introduction}

Under the generic title of ``Gibbs Paradox'' one usually considers a number
of related questions in both phenomenological thermodynamics and in
statistical mechanics: (1) The entropy change when two distinct gases are
mixed happens to be independent of the nature of the gases; is this in
conflict with the idea that in the limit as the two gases become identical
the entropy change should vanish? (2) Should the thermodynamic entropy of
Clausius be extensive or not? Is this a mere convention, or a hard
experimental fact? (3) Should two microstates that differ only in the
exchange of identical particles be counted as two or just one microstate?

The conventional wisdom, as recorded in many textbooks, (see \emph{e.g.} 
\cite{Callen85}\cite{Hestenes70}) asserts that the resolution of the paradox
rests on quantum mechanics. This analysis is unsatisfactory on many counts;
at best it is incomplete. While it is true that the exchange of identical
quantum particles does not lead to a new microstate this approach ignores
the case of classical, and even non-identical particles. Nanoparticles in a
colloidal suspension or macromolecules in solution provide us with simple
examples:\ should the entropy of such systems of non-identical classical
particles be extensive or not?

A number of authors (including Grad, van Kampen, and Jaynes among others)
have recognized that quantum theory has no bearing on the matter. Grad \cite
{Grad61} and van Kampen \cite{vanKampen84} approach the problem from both
the point of view of phenomenological classical thermodynamics and of
statistical mechanics. Theirs is the orthodox version of a statistical
mechanics that ultimately rests upon ergodic theory. Jaynes \cite{Jaynes92}
addresses the subject as a problem in phenomenological thermodynamics (see
also \cite{Tribus61}). Perhaps surprisingly, Jaynes does not provide a
statistical analysis from the point of view of information theory; it is
only in the last few paragraphs of \cite{Jaynes92}, and then only as a
speculation, that he suggests that a more fundamental statistical treatment
rooted in information theory might provide a more satisfactory and complete
analysis.

Whether one accepts information theory or ergodic theory as the correct
foundation for statistical mechanics turns out to be immaterial in this
case: all three of those authors agree that the paradox is resolved once it
is realized that (1) the experimental data is usually silent on the matter
of whether the thermodynamic entropy is extensive or not (extensivity is no
more than a convenient convention). (2) There is no such thing as \emph{the}
entropy of a system, that there are \emph{many} entropies. And (3) the
appropriate choice of entropy, or equivalently the choice between treating
particles as being distinguishable or not depends on the resolution of the
particular experiment being performed.

The purpose of this paper is essentially pedagogical; we add to their
analysis by examining the paradox from the point of view of information
theory. Our argument is guided by a certain property of entropy, the
`grouping' property, that Shannon recognized as an essential requirement on
any measure of information. (Briefly, this property is the statement that
the entropy of a probability distribution over a set of alternatives is
unchanged if the alternatives are grouped into subsets, each with its own
entropy.) This provides the right language, and therefore a useful guide in
navigating past various conceptual difficulties. For example, the usual
treatments of thermodynamics might mislead one to think in terms of only one
entropy. On the other hand, the usual treatments of statistical mechanics,
aim at providing the link between descriptions in terms of `micro' and
`macro' states, and in the process, might mislead one to presuppose that no
other intermediate `meso' states are relevant. Focusing on the `grouping'
property steers the discussion in the right direction, namely towards
studying the connection between the different entropies corresponding to
different descriptions of the same system.

This paper is organized as follows: In section 2 we review the `grouping'
property and in section 3 we specify what we mean by the various states that
enter our discussion. Sections 4 and 5 discuss the cases of identical and of
non identical particles. In section 6 we offer some final remarks.

\section{The `grouping' property}

Perhaps the easiest way to enunciate the `grouping' property is to prove it.

Our choice of notation and language will reflect, from the very start, the
fact that our subject is statistical mechanics. A physical system can be in
any one of a set of alternative states which will be labeled $i=1,2,\ldots $%
. Knowing that the system is found in state $i$ requires complete
information up to the most minute details about every particle in the
system, their location, momentum, internal state, etc. Such detailed states
are called microstates. For a classical system the microstates should more
properly be described as points in a phase space continuum. It is
convenient, however, to divide phase space into cells of arbitrarily small
size and consider discrete classical microstates.

The precise microstate of the system is not known; the limited information
we possess allows us at best to assign probability $p_i$ to each microstate $%
i$. The amount of additional information that would allow us to pinpoint the
actual microstate is given by the entropy of the distribution $p_i$, 
\begin{equation}
S[p]=-\sum_i\,p_i\log p_i\,.  \label{S[p]}
\end{equation}
The set of microstates $i$ (phase space) can be partitioned into
non-overlapping subsets; the subsets are groups of microstates which will be
labeled $g$. The sum in eq.(\ref{S[p]}) can then be rearranged into 
\begin{equation}
S[p]=-\sum_g\sum_{i\in g}\,p_i\log p_i\,.  \label{S1[p]}
\end{equation}
The probability that the system is found in group $g$ be 
\begin{equation}
P_g=\sum_{i\in g}\,p_i\,.  \label{Pg}
\end{equation}
Next consider the sum over groups $g$ in eq.(\ref{S1[p]}). Each term in this
sum can be rearranged as follows 
\begin{equation}
-\sum_{i\in g}\,p_i\log p_i=-P_g\sum_{i\in g}\,\frac{p_i}{P_g}\log \frac{p_i%
}{P_g}-P_g\log P_g\,.
\end{equation}
Let $p_{i|g}$ denote the conditional probability that the system is in
microstate $i\in g$ given we know it is in the group $g$, 
\begin{equation}
p_{i|g}=\frac{p_i}{P_g}\quad .  \label{pi|g}
\end{equation}
Then eq.(\ref{S[p]}) can be written as 
\begin{equation}
S=S_G+\sum_gP_gS_g\,,  \label{grouping prop}
\end{equation}
where 
\begin{equation}
S_G=-\sum_gP_g\log P_g  \label{SG}
\end{equation}
and 
\begin{equation}
S_g=-\sum_{i\in g}\,p_{i|g}\log p_{i|g}\,.
\end{equation}

Eq.(\ref{grouping prop}) is the `grouping' property we seek. It can be read
as follows: the information required to locate the system in one of its
microstates $i$ equals the information required to locate the system in one
of the groups $g$, plus the expectation over the groups of the information
required to pinpoint the microstate within each group.

Shannon \cite{Shannon48} regarded the `grouping' property as an obvious
requirement to be imposed on any measure of information and included it
among his axioms leading to eq.(\ref{S[p]}). To others, who did not seek a
measure of amount information, but rather a method to take information into
account, eq.(\ref{grouping prop}) is not an obviously unavoidable
requirement. Thus, other derivations of eq.(\ref{S[p]}) have been proposed 
\cite{ShoreJohnson80} from which entropy emerges as a tool for consistent
reasoning. In these approaches entropy does not need to be interpreted in
terms of heat, disorder, uncertainty, or lack of information: entropy needs
no interpretation.

\section{Entropies and descriptions}

The `grouping' property, eq.(\ref{grouping prop}), plays an important role
because it establishes a relation between two different descriptions and, in
doing so, it invokes three different entropies (none of which is the
thermodynamical entropy of Clausius).

We describe the system with high resolution as being in any one of its
microstates $i$ with probability $p_i$, or alternatively, with lower
resolution as being in any one of the groups $g$ with probability $P_g$.
Since the description in terms of groups is less detailed we might refer to
them as `mesostates'.

A thermodynamical description, on the other hand, corresponds to neither the
high resolution description in terms of microstates, nor the lower
resolution description in terms of mesostates. It is a description that
incorporates the least amount of information necessary for a totally
macroscopic characterization of equilibrium. The state that is relevant here
is defined by the values of those variables the control of which guarantees
the macroscopic reproducibility of experiments. Such states are called
macrostates. The typical thermodynamic variables include the energy, volume,
magnetization, etc., but here, for simplicity, we will consider only the
energy since including other macrovariables is straightforward and does not
modify the gist of the argument.

The standard connection between the thermodynamic description in terms of
macrostates and the description in terms of microstates is established using
the Method of Maximum Entropy. Let the energy of microstate $i$ be $%
\varepsilon _i$. To the macrostate of energy $E$ we associate that
probability distribution $p_i$ which maximizes the entropy (\ref{S[p]})
subject to the constraints 
\begin{equation}
\sum_i\,p_i=1\quad \text{and}\quad \sum_i\,p_i\varepsilon _i=E\,.
\end{equation}
The well-known result is the canonical distribution, 
\begin{equation}
p_i=\frac{e^{-\beta \varepsilon _i}}{Z_H}\,,  \label{can dist H}
\end{equation}
where the partition function $Z_H$ and the Lagrange multiplier $\beta $ are
determined from 
\begin{equation}
Z_H=\sum_ie^{-\beta \varepsilon _i}\quad \text{and\quad }\frac{\partial \log
Z_H}{\partial \beta }=-E\,.
\end{equation}
The corresponding entropy, obtained by substituting eq.(\ref{can dist H})
into eq.(\ref{S[p]}), 
\begin{equation}
S_H=\beta E+\log Z_H\,,
\end{equation}
measures the amount of information beyond the value $E$ to specify the
microstate.

Before we compute and interpret the probability distribution over mesostates
and its corresponding entropy we must be more specific about which
mesostates we are talking about. This is what we do next.

\section{Identical particles}

Consider a system of $N$ classical particles that are exactly identical. The
particles may or may not interact with each other (\emph{i.e.}, the argument
is not limited to ideal gases).

The interesting question is whether these identical particles are also
`distinguishable'. By this we mean the following: we look at two particles
now and we label them. We look at the particles later; somebody might have
switched them. Can we tell which particle is which? The answer is: it
depends. Whether we can distinguish identical particles or not depends on
whether we were able (and willing) to follow their trajectories.

A slightly different version of the same question concerns an $N$-particle
system in a certain state. Some particles are permuted. Does this give us a
different state? As discussed in the previous section, the answer to this
question requires a careful specification of what we mean by a state.

If by state we mean a microstate, that is a point in the $N$-particle phase
space, then a permutation does indeed lead to a new microstate. On the other
hand, our concern with permutations suggests that it is useful to introduce
the notion of a mesostate defined as the group of those $N!$ microstates
that are obtained as permutations of each other. With this definition it is
clear that a permutation of the identical particles does not lead to a new
mesostate.

Now we can return to discussing the connection between the thermodynamic
macrostate description and the description in terms of mesostates using, as
before, the Method of Maximum Entropy. Since the particles are identical,
all those microstates $i$ within the same mesostate $g$ have the same
energy, which we will denote by $\varepsilon _g$ (\emph{i.e.}, $\varepsilon
_i=\varepsilon _g$ for all $i\in g$). To the macrostate of energy $E$ we
associate that probability distribution $P_g$ which maximizes the entropy (%
\ref{SG}) subject to the constraints 
\begin{equation}
\sum_g\,P_g=1\quad \text{and}\quad \sum_g\,P_g\varepsilon _g=E\,.
\end{equation}
The result is also a canonical distribution, 
\begin{equation}
P_g=\frac{e^{-\beta \varepsilon _g}}{Z_L}\,,  \label{can dist L}
\end{equation}
where 
\begin{equation}
Z_L=\sum_ge^{-\beta \varepsilon _g}\quad \text{and\quad }\frac{\partial \log
Z_L}{\partial \beta }=-E\,.
\end{equation}
The corresponding entropy, obtained by substituting eq.(\ref{can dist L})
into eq.(\ref{SG}), 
\begin{equation}
S_L=\beta E+\log Z_L\,,
\end{equation}
measures the amount of information beyond the value $E$ to specify the
mesostate.

Notice that two different entropies $S_H$ and $S_L$ have been assigned to
the same macrostate $E$; they measure the different amounts of additional
information required to specify the state of the system to a high resolution
(the microstate) or to a low resolution (the mesostate).

The relation between $Z_H$ and $Z_L$ can be obtained from eqs.(\ref{Pg}), (%
\ref{can dist H}) and (\ref{can dist L}): 
\begin{equation}
Z_L=\frac{Z_H}{N!}\,.  \label{ZLZH}
\end{equation}
The relation between $S_H$ and $S_L$ is obtained from the `grouping'
property. First use eq.(\ref{pi|g}) to get $p_{i|g}=1/N!$, and then
substitute into eq.(\ref{SG}) (with $S=S_H$ and $S_G=S_L$) to get 
\begin{equation}
S_L=S_H-\log N!\,.  \label{SLSH}
\end{equation}

Equations (\ref{ZLZH}) and (\ref{SLSH}) both exhibit the Gibbs $N!$
`corrections'. Our analysis shows (1) that the justification of the $N!$
factor is not to be found in quantum mechanics, and (2) that the $N!$ does
not correct anything. The $N!$ is not a fudge factor that fixes a wrong
(possibly nonextensive) entropy $S_H$ into a correct (possibly extensive)
entropy $S_L$. Both entropies $S_H$ and $S_L$ are correct. They differ
because they measure different things: one measures the information to
specify the microstate, the other measures the information to specify the
mesostate.

An important goal of statistical mechanics is to provide a justification, an
explanation of thermodynamics. Thus, we still need to ask which of the two
statistical entropies, $S_H$ or $S_L$, should be identified with the
thermodynamical entropy of Clausius $S_{\text{exp}}$. Inspection of eqs.(\ref
{ZLZH}) and (\ref{SLSH}) shows that, as long as one is not concerned with
experiments that involve changes in the number of particles, the same
thermodynamics will follow whether we set $S_H=S_{\text{exp}}$ or $S_L=S_{%
\text{exp}}$. This is the conclusion reached by Grad, van Kampen and Jaynes.

But, of course, experiments involving changes in $N$ are very important (for
example, in the equilibrium between different phases, or in chemical
reactions). Since in the usual thermodynamical experiments we only care that
some number of particles has been exchanged, and we do not care which were
the actual particles exchanged, we expect that the correct identification is 
$S_L=S_{\text{exp}}$. Indeed, the quantity that regulates the equilibrium
under exchanges of particles is the chemical potential defined by 
\begin{equation}
\mu _{\text{exp}}=-kT\left( \frac{\partial S_{\text{exp}}}{\partial N}%
\right) _{E,V,\ldots }
\end{equation}
(This is analogous to the temperature, an intensive quantity that regulates
the equilibrium under exchanges of heat.) The two identifications $S_H=S_{%
\text{exp}}$ or $S_L=S_{\text{exp}}$, lead to two different chemical
potentials, related by 
\begin{equation}
\mu _L=\mu _H-NkT\,.
\end{equation}
It is easy to verify that, under the usual circumstances where surface
effects can be neglected relative to the bulk, $\mu _L$ has the correct
functional dependence on $N$: it is intensive and can be identified with $%
\mu _{\text{exp}}$. On the other hand, $\mu _H$ is not an intensive quantity
and cannot therefore be identified with $\mu _{\text{exp}}$.

\section{Non-identical particles}

In the last section we saw that classical identical particles can be
treated, depending on the resolution of the experiment, as being
distinguishable or indistinguishable. In this section we go further and
point out that even non-identical particles can be treated as
indistinguishable. Our goal is to state explicitly in precisely what sense
it is up to the observer to decide whether particles are distinguishable or
not.

We defined a mesostate as a subset of $N!$ microstates that are obtained as
permutations of each other. With this definition it is clear that a
permutation of particles does not lead to a new mesostate even if the
exchanged particles are not identical. This is an important extension
because, unlike quantum particles, classical particles cannot be expected to
be exactly identical down to every minute detail. In fact in many cases they
are grossly different. Consider the example of milk, \emph{i.e.}, a
colloidal suspension of fat droplets in water, or a solution of
macromolecules. A high resolution device, for example an electron
microscope, would reveal that no two fat droplets or two macromolecules are
exactly alike. And yet, for the purpose of modelling most of our macroscopic
observations (\emph{i.e.}, the thermodynamics of milk) it is not necessary
to take account of the myriad ways in which two fat droplets can differ.

Consider a system of $N$ particles. We can perform rather crude macroscopic
experiments the results of which can be summarized with a simple
phenomenological thermodynamics where $N$ is one of the relevant variables
that define the macrostate. Our goal is to construct a statistical
foundation that will explain this macroscopic model, reduce it, so to speak,
to `first principles'. The particles might ultimately be non-identical, but
the crude phenomenology is not sensitive to their differences and can be
explained by postulating mesostates $g$ and microstates $i$ with well
defined energies $\varepsilon _i=\varepsilon _g$, for all $i\in g$, as if
the particles were identical. As in the previous section this statistical
model gives

\begin{equation}
Z_L=\frac{Z_H}{N!}\quad \text{with\quad }Z_H=\sum_ie^{-\beta \varepsilon
_i}\,,
\end{equation}
and the connection to the thermodynamics is established by postulating 
\begin{equation}
S_{\text{exp}}=S_L=S_H-\log N!\,.
\end{equation}

Next we consider what happens when more sophisticated experiments are
performed. The examples traditionally offered in discussions of this sort
refer to the new experiments made possible by the discovery of membranes
that are permeable to some of the $N$ particles but not to the others.
Other, perhaps historically more realistic examples, are afforded by the
availability of new experimental data, for example, more precise
measurements of a heat capacity as a function of temperature, or perhaps
measurements in a range of temperatures that had previously been
inaccessible.

Suppose the new phenomenology can be modelled by postulating the existence
of two kinds of particles. (Experiments that are even more sophisticated
might allow us to detect three or more kinds, perhaps even a continuum of
different particles.) What we previously thought were $N$ identical
particles we will now think as being $N_a$ particles of type $a$ and $N_b$
particles of type $b$. The new description is in terms of macrostates
defined by $N_a$ and $N_b$ as the relevant variables.

To construct a statistical explanation of the new phenomenology from `first
principles' we need to revise our notion of mesostate. Each new mesostate
will be a group of microstates which will include all those microstates
obtained by permuting the $a$ particles among themselves, and by permuting
the $b$ particles among themselves, but will not include those microstates
obtained by permuting $a$ particles with $b$ particles. The new mesostates,
which we will label $\hat{g}$ and to which we will assign energy $%
\varepsilon _{\hat{g}}$, will be composed of $N_a!N_b!$ microstates $\hat{%
\imath}$, each with a well defined energy $\varepsilon _{\hat{\imath}%
}=\varepsilon _{\hat{g}}$, for all $\hat{\imath}\in \hat{g}$. The new
statistical model gives

\begin{equation}
\hat{Z}_L=\frac{\hat{Z}_H}{N_a!N_b!}\quad \text{with\quad }\hat{Z}_H=\sum_{%
\hat{\imath}}e^{-\beta \varepsilon _{\hat{\imath}}}\,,
\end{equation}
and the connection to the new phenomenology is established by postulating 
\begin{equation}
S_{\text{new exp}}=\hat{S}_L=\hat{S}_H-\log N_a!N_b!\,.
\end{equation}

In discussions of this topic it is not unusual to find comments to the
effect that in the limit as particles $a$ and $b$ become identical one
expects that the entropy of the system with two kinds of particles tends to
the entropy of a system with just one kind of particle. The fact that this
expectation is not met is one manifestation of the Gibbs paradox.

From the information theory point of view the paradox does not arise because
there is no such thing as \emph{the entropy of the system}, there are
several entropies. It is true that as $a\rightarrow b$ we will have $\hat{Z}%
_H\rightarrow Z_H$, and accordingly $\hat{S}_H\rightarrow S_H$, but there is
no reason to expect a similar relation between $\hat{S}_L$ and $S_L$ because
these two entropies refer to mesostates $\hat{g}$ and $g$ that remain
different even as $a$ and $b$ became identical. In this limit the mesostates 
$\hat{g}$, which are useful for descriptions that treat particles $a$ and $b$
as indistinguishable among themselves but distinguishable from each other,
lose their usefulness.

\section{Conclusion}

We conclude with a comment and a quotation. First, our comment.

The Gibbs paradox in its various forms arises from the widespread
misconception that entropy is a real physical quantity and that one is
justified in talking about \emph{the entropy of the system}. The
thermodynamic entropy is not a property of the system. It is somewhat more
accurate to assert that entropy is a property of our description of the
system, it is a property of the macrostate. More explicitly, it is a
function of the macroscopic variables used to define the macrostate. To
different macrostates reflecting different choices of variables there
correspond different entropies for the very same system.

But this is not the complete story: the entropy is not just a function of
the macrostate. Entropies reflect a relation between two descriptions of the
same system: in addition to the macrostate, we must also specify the set of
microstates, or the set of mesostates, as the case might be. Then, having
specified the macrostate, an entropy can be interpreted as the amount of
additional information required to specify the microstate or mesostate. We
have found the `grouping' property very valuable precisely because it
emphasizes this dependence of entropy on a second argument, namely, the
choice of micro or mesostates.

The promissed quotation is a remark by van Kampen \cite{vanKampen84} that
very aptly captures the issue of whether identical classical particles are
distinguishable or not:

\begin{quotation}
``The question is not whether the particles are identical in eyes of God,
but merely in the eyes of the beholder.''
\end{quotation}

\noindent This is a surprisingly perceptive remark, particularly coming from
someone who strongly opposed the information theory approach to statistical
mechanics.


\begin{thebibliography}{9}
\bibitem{Callen85}  L. D. Landau and E. M. Lifshitz, \emph{Statistical
Physics} (Pergamon, New York, 1977); H. B. Callen, \emph{Thermodynamics and
an Introduction to Thermostatistics} (Wiley, New York, 1985).

\bibitem{Hestenes70}  For a wider variety of views and comments on this
subject see \emph{e.g.}:\emph{\ }Hestenes, D., Am. J. Phys. \textbf{38}, 840
(1970); Boyer, T. H., Am. J. Phys. \textbf{38}, 849 (1970); Dicks, D. and
van Dij, V., Am. J. Phys. \textbf{56}, 430 (1988); P. D. Pesic, Am. J. Phys. 
\textbf{59}, 971 (1991); Baierlein R., Am. J. Phys. \textbf{65}, 314 (1997).

\bibitem{Grad61}  H. Grad, ``The Many Faces of Entropy,'' Comm. Pure and
Appl. Math. \textbf{14}, 323 (1961), and ``Levels of Description in
Statistical Mechanics and Thermodynamics'' in \emph{Delaware Seminar in the
Foundations of Physics}, ed. by M. Bunge (Springer-Verlag, New York, 1967).

\bibitem{vanKampen84}  N. G. van Kampen, ``The Gibbs Paradox'' in \emph{%
Essays in Theoretical Physics in Honor of Dirk ter Haar}, ed. by W. E. Parry
(Pergamon, Oxford, 1984).

\bibitem{Jaynes92}  E. T. Jaynes, ``The Gibbs Paradox'' in \emph{Maximum
Entropy and Bayesian Methods}, ed. by C. R. Smith, G. J. Erickson and P. O.
Neudorfer (Kluwer, Dordrecht, 1992).

\bibitem{Tribus61}  M. Tribus, \emph{Thermostatics and Thermodynamics} (van
Nostrand, New York, 1961) also discusses the issue of distinguishability
from a point of view that is motivated by information theory.

\bibitem{Shannon48}  C. E. Shannon, Bell Systems Tech. Jour. \textbf{27},
379, 623 (1948); C. E. Shannon and W. Weaver, \emph{The Mathematical Theory
of Communication} (Univ. of Illinois Press, Urbana, 1949).

\bibitem{ShoreJohnson80}  J. E. Shore and R. W. Johnson, ``Axiomatic
derivation of the Principle of Maximum Entropy and the Principle of Minimum
Cross-Entropy,'' IEEE Trans. Inf. Theory \textbf{IT-26}, 26 (1980); Y.
Tikochinsky, N. Z. Tishby and R. D. Levine, Phys. Rev. Lett. \textbf{52},
1357 (1984) and Phys. Rev. \textbf{A30}, 2638 (1984); J. Skilling, ``The
Axioms of Maximum Entropy'' in \emph{Maximum-Entropy and Bayesian Methods in
Science and Engineering}, G. J. Erickson and C. R. Smith (eds.) (Kluwer,
Dordrecht, 1988).
\end{thebibliography}
\end{document}